\documentclass[useAMS,usenatbib]{mn2e}
\usepackage{times}
\usepackage{rotating}
\usepackage{amsmath}
\usepackage{amssymb}
\usepackage[dvipsnames]{xcolor}
\usepackage{graphicx}
\usepackage{graphics}
\usepackage{hyperref}
\hypersetup{
     colorlinks   = true,
     citecolor    = blue
}

\long\def\symbolfootnote[#1]#2{\begingroup%
\def\thefootnote{\fnsymbol{footnote}}\footnote[#1]{#2}\endgroup}
\newcommand{\gae}{\lower 2pt \hbox{$\, \buildrel {\scriptstyle >}\over {\scriptstyle
\sim}\,$}}
\newcommand{\lae}{\lower 2pt \hbox{$\, \buildrel {\scriptstyle <}\over {\scriptstyle
\sim}\,$}}

\def\d{\mathrm{d}}

\def\nGJ{\mbox{\large n}_{_{GJ}}}

\def\bk{{\bf k}}

\def\bkp{{\bf k'}}
\def\bkps{{\bf k_1'}}
\def\hatbkp{\hat{\bf k}'}
\def\hatbkps{\hat{\bf k}_1'}
\def\nk{n_{\bf k}}

\def\nkp{n_{\bf k'}}
\def\nkps{n_{\bf k_1'}}

\begin{document}

\title[Radiation forces and FRBs]
{Radiation Forces Constrain the FRB Mechanism} 

\author[Kumar \& Lu]{Pawan Kumar$^{1}$\thanks{pk@astro.as.utexas.edu} and Wenbin
  Lu$^{2}$\thanks{wenbinlu@caltech.edu} 
\\ $^{1}$Department of Astronomy, University of Texas at Austin, Austin,
 TX 78712, USA\\
$^{2}$TAPIR, Walter Burke Institute for Theoretical Physics, Mail Code
350-17, Caltech, Pasadena, CA 91125, USA 
}

% \pagerange{\pageref{000}--\pageref{000}} \pubyear{2017}

\maketitle

\begin{abstract}

We provide constraints on Fast Radio Burst (FRB) models
by careful considerations of radiation forces associated with these powerful 
transients. We find that the induced-Compton scatterings of the coherent 
radiation by electrons/positrons accelerate particles to very large
Lorentz factors (LF) in and around the source of this radiation. This 
severely restricts those models for FRBs that invoke relativistic shocks 
and maser type instabilities at distances less than about 10$^{13}$cm of the 
neutron star. Radiation traveling upstream, in these
models, forces particles to move away from the shock with a LF larger 
than the LF of the shock front. This suspends the photon generation 
process after it has been operating for less than ~$\sim 0.1$ ms (observer 
frame duration). We show that masers operating in shocks at 
distances larger than 10$^{13}$cm cannot simultaneously account for the 
burst duration of 1 ms or more and the observed $\sim$GHz frequencies of 
FRBs without requiring an excessive energy budget ($>10^{46}$erg); 
the energy is not calculated by imposing any efficiency consideration, or 
other details, for the maser mechanism, but is entirely the result of 
ensuring that particle acceleration by induced-Compton forces upstream
of the shock front does not choke off the maser process.
 For the source to operate more or less continuously for a few ms, 
it should be embedded in a strong magnetic field -- cyclotron frequency $\gg$ 
wave frequency -- so that radiation forces don't disperse the plasma and 
shut-off the engine.

\end{abstract}


\section{Introduction}

Fast radio bursts (FRBs) are milli-second-duration bright (flux 
density $\sim$ Jansky) transient events observed 
between 400 MHz and 7 GHz 
frequencies (Amiri et al. 2019a, 2019b; Ravi 2019a, 2019b; Ravi et al. 2019;
Oslowski et al. 2019; Kocz et al. 2019; Bannister et al. 2019; 
Gajjar et al. 2018; Michilli et al. 2018; Farah et al. 2018; 
Shannon et al. 2018; Bannister et al. 2017; Law et al. 2017;
Chatterjee et al. 2017; Marcote et al. 2017; Tendulkar et al. 2017;
Spitler et al. 2016; Petroff et al. 2016; Spitler et al. 2014; 
Thornton et al. 2013; Lorimer et al. 2007).
Numerous mechanisms have been suggested for the generation of the coherent 
radio emission of FRBs, eg. Kumar et al. (2017), Metzger et
al. (2017), Yang \& Zhang (2018), Lu \& Kumar (2018), Metzger et al. (2019);
for recent reviews see Katz 
(2018), Petroff et al. (2019), Codes \& Chatterjee (2019). 
Most of these mechanisms
are proposed to operate at a distance from neutron star surface of 
10$^{13}$ cm or less. We explore in this paper constraints on the FRB
radiation mechanisms provided by rapid acceleration of charge particles 
by the strong electric field associated with the radiation and induced 
Compton scatterings of photons; induced
Compton scattering refers to the scattering of a photons by electrons when
the occupation number of photon quantum states is much larger than unity
(for FRBs the occupation number is of order 10$^{35}$). The electron-photon 
scattering optical depth is increased due to the induced-Compton effect, 
which has been calculated by many authors, e.g. Melrose (1971), 
Blandford (1973), Blandford \& Scharlemann (1975), Wilson \& Rees (1978). 
We include the constraint on FRB models provided by induced-Compton (IC) 
optical depth, but that turns out to much weaker than the IC acceleration
that is estimated here. Particle acceleration
due to the electric field of FRB radiations and its implications are 
considered in \S2. Acceleration due to induced-Compton scatterings is
described in \S3, and the constraints these processes and some other
general considerations impose on the FRB source and radiation mechanisms
is discussed in \S4.

\section{Particle motion due to large amplitude EM wave and static magnetic 
field}
\label{pmEMB}

The RMS electric field strength associated with FRB radiation at a distance 
$R$ from the source is
\begin{equation}
   {E_0\over \sqrt{2}} = \left({L\over c R^2}\right)^{1/2} = (1.8{\rm x}10^3\, 
       {\rm esu}) L_{43}^{1/2} R_{13}^{-1},
\end{equation}
where $L$ is isotropic equivalent luminosity of the FRB. The non-linearity 
parameter, $a$, associated with this field -- which is a rough measure
of the energy gained by an electron traveling a distance of 
one wavelength in the wave-electric-field divided by its rest mass-energy -- is
\begin{equation}
  a \equiv {q E_0\over m c \omega} = 4.5\, L_{43}^{1/2} R_{13}^{-1} 
         \omega_{10}^{-1},
   \label{aparam}
\end{equation}
where $q$ and $m$ are electron charge and mass respectively, and $\omega$
is the wave frequency (radian s$^{-1}$). 

Let us consider the motion of a particle exposed to a linearly polarized EM wave
and a uniform magnetic field that is perpendicular to the wave-vector of the
EM wave. The wave EM fields and vector potentials for the EM wave and 
the static magnetic field are as follows:

\begin{align}
    {\bf E_w} & = E_0 {\bf \hat{x}} \sin(kz - \omega t), \quad {\bf B_w}
     = E_0 {\bf \hat{y}} \sin(kz - \omega t), \\
    {\bf A_w} & = -{cE_0\over \omega} {\bf \hat{x}} \cos(kz - \omega t), 
     \quad {\bf A_B} = (B_y {\bf \hat{x}} -B_x {\bf \hat{y}}) z.
\end{align}

The Lagrangian for particle motion is
\begin{equation}
   L = -{m c^2\over\gamma} + {q\over c} {\bf A\cdot v},
\end{equation}
where $\gamma$ is the LF of the particle, ${\bf v}$ is its
3-velocity, and ${\bf A} = {\bf A_w} + {\bf A_B}$ is the vector potential
for the EM wave plus the static magnetic field. The $x$ and $y$ components 
of the particle's canonical momentum are conserved since the Lagrangian is
independent of $x$ and $y$ coordinates--
\begin{equation}
   p_x = m \gamma v_x + {q A_x\over c} = m\gamma v_x - {qE_0\over\omega}
        \cos\psi + {q B_y z\over c} = constant,
 \label{px}
\end{equation}
\begin{equation}
  p_y = m\gamma v_y - {q B_x z\over c} = constant,
 \label{py}
\end{equation}
where
\begin{equation}
   \psi = kz - \omega t.
\end{equation}
The $z$-component of the momentum equation 
\begin{equation}
   {d \gamma v_z\over dt} = {q\over mc} \left[ B_w v_x + v_x B_y 
      - v_y B_x\right],
\end{equation}
can be rewritten as
\begin{equation}
  {d\over dt}\left[\gamma v_z - \gamma c - \omega_B(x\sin\theta_B - 
       y\cos\theta_B)\right] = 0, 
 \label{pz}
\end{equation}
where
\begin{equation}
   \sin\theta_B = B_y/B, \quad{\rm and}\quad\omega_B = {q B\over mc}
\end{equation}
is the cyclotron frequency.

We assume that the particle is initially at rest, i.e. ${\bf v = 0}$, before it 
is hit by the EM wave, and that its initial position is ${\bf r = 0}$.
Thereafter its velocity is obtained from equations 
(\ref{px}), (\ref{py}) and (\ref{pz}) by applying the initial conditions,
\begin{align}
    \gamma v_x & = ac\left[ \cos(kz - \omega t) - 1\right] - 
       \omega_B z \sin\theta_B, \\
    \gamma v_y & = \omega_B z \cos\theta_B, \\
    \gamma v_z & = c(\gamma-1) + \omega_B(x \sin\theta_B - y \cos\theta_B),
\end{align}
where $a$ is given by equation (\ref{aparam}).

The particle LF is easily obtained from these equations and is
\begin{multline}
    \gamma = {4a^2 \sin^4(\psi/2) + 4a\sin^2(\psi/2)\xi_z\sin\theta_B 
     + \xi_z^2 + 1\over 2[1 - \xi_\rho\sin(\theta_B-\phi)]}  \\
       \quad\quad + [1 - \xi_\rho \sin(\theta_B-\phi)]/2
   \label{gam_E1}
\end{multline}
where 
\begin{equation}
   \sin\phi = y/\rho, \quad \rho^2 = x^2 + y^2,
        \quad \xi_\rho = \rho\omega_B/c, \quad \xi_z = z\omega_B/c.
\end{equation}

\begin{figure}
\centerline{\hbox{\includegraphics[width=9cm, height=13cm, angle=0]{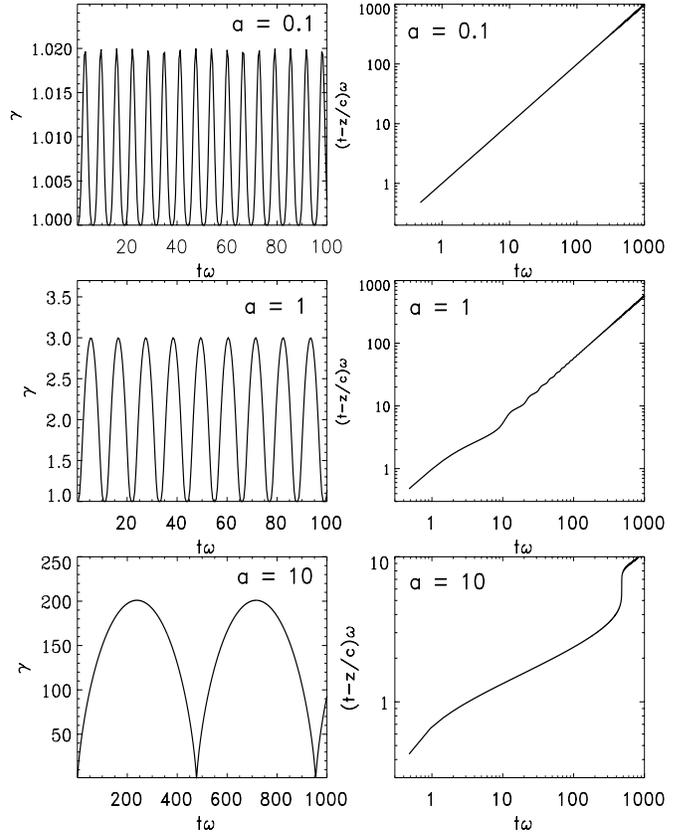}}}
\vskip -0.7cm
\caption{Shown on the left panels are the particle LF, $\gamma$, 
as a function of dimensionless time, $t\omega$, for three different values
of $a$: 0.1, 1 and 10 (top to bottom); where $\omega$ is the EM wave
frequency. These calculations are for weak magnetic fields such that
the ratio of the electron cyclotron frequency and the EM wave frequency 
($\omega_B/\omega$) is 10$^{-5}$; results for $\omega_B/\omega=10^{-3}$
are effectively the same as shown here. The three right panels show the
location of the particle wrt the front of the EM wave, i.e. $[t - z(t)/c]
\omega$; for large values of $a$, the particle rides the wave at a roughly
constant phase angle for a long duration of time (see the bottom right 
panel for $a=10$).
}
\label{fig1}
\end{figure}

The particular case of $\omega_B=0$, i.e. vanishing static magnetic field, 
is illuminating:
\begin{equation}
    \gamma v_x = -2ac \sin^2(\psi/2), \; \gamma v_z = c(\gamma-1), 
    \; \gamma = 1 + 2a^2\sin^4(\psi/2).
\end{equation}
For $a>1$, the particle LF is of order $2 a^2$ and its velocity 
vector lies within an angle $\sim a^{-1}$ of the wave propagation direction. 
For $a\ll 1$, the velocity component along the wave vector oscillates at 
frequency $2\omega$ and perpendicular to it at frequency $\omega$. However,
for $a>1$, the phase function along the particle worldline $|\psi| = 
\omega t - kz(t) \sim \omega t/2\langle
\gamma^2 \rangle$, and so the particle LF oscillates at frequency $\sim
\omega/\langle \gamma^2 \rangle \sim \omega/[2(1+a^2)]$; where $\langle
\gamma^2 \rangle$ is the time averaged LF-squared of the particle.
Figure  \ref{fig1} shows numerical results for particle trajectories for a
few different values of $a$.

Particles out to a very large distance from the source of the intense
FRB radiation are vigorously agitated and forced to move at high velocities -- 
the time averaged LF $\gamma\sim a^2\sim 10^2 R_{13}^{-2}$; the nonlinearity 
parameter for those FRBs which emit a good fraction of their energy at 
$\sim 500$ MHz frequency is: $a \sim 10\, R_{13}^{-1} L_{43}^{1/2}$ (see eq. 
\ref{aparam}).

Considering that charge particles are forced to move at close to the speed
of light by the electric field of the radiation out to a distance 
$\sim 10^{14}$cm from the source, it might be hard for the plasma lens 
model (Cordes et al. 2017; Main et al. 2018) to operate within this 
distance of the source -- a few ms duration radiation pulse cannot
pass through the plasma in the ``lens'' if the plasma is located at a
distance $\lae 10^{13}$cm from the FRB source since the plasma would be
forced to move at speed $\sim c$ by the wave electric field -- unless the 
magnification factor of the plasma lens is much larger than $\sim 10^2$.

If the FRB radiation is produced in shock heated plasma (e.g. 
Metzger et al. 2019) at radius $R$, then particles upstream of 
the shock front would be accelerated to LF $\gamma\approx
2 a^2 \sim 50 L_{43} R_{13}^{-2} \omega_{10}^{-2}$ (eq. \ref{aparam}) by the
electric field of radiation moving upstream as long as the composition of the
medium is $e^\pm$ and the cyclotron frequency is much less than 1 GHz.
The LF of upstream particles wrt to the shock in this case is 
$\sim \gamma_{sf}/2\gamma$; $\gamma_{sf}$ is the LF of the shock front 
wrt to the undisturbed, upstream, medium. The upstream medium is also 
compressed by a factor $\sim 4 a^2$ due to this acceleration, and the
component of upstream particle four-velocity tangential to the shock-surface
is $\sim a$ which varies on a time scale of the wave period. These effects
would modify the growth rate of synchrotron maser instability operating 
near the shock front, and reduce the efficiency of converting blast wave
energy to coherent GHz radiation.
 
Furthermore, particle acceleration upstream of the shock front would shut 
off radiation production for a while in the observed band, if $R\lae 10^{13}$ 
cm, until upstream particles slow down by plowing into the plasma further out.
So the radiation by shocked plasma is not going to be produced continuously
if $R\lae 10^{13}$cm.

Another effect of particle acceleration by the intense FRB radiation is
that it depletes the energy from the the outward moving radiation front --
charge particles undergoing acceleration radiate -- at a rate much larger
than one might expect from Thomson scatterings. This is estimated
in the following sub-section, where we also discuss the constraints 
imposed by this process on the circum-burst medium.  

\subsection{Radiative loss of Wave-accelerated particles}

We calculate in this sub-section the energy loss suffered by FRB pulse
as it travels through the CSM of the magnetar due to power radiated by 
particles that are accelerated by the EM field of the FRB radiation (particle
acceleration was calculated in \S2).
The power emitted by a relativistic particle undergoing acceleration 
is given by (the covariant form of) the Larmor formula
\begin{equation}
  \label{eq:1}
  P = {2q^2 \over 3c}\left[-\left(d u^0\over d \tau\right)^2 + \sum_{i=x, y,
      z} \left(d u^i\over d \tau\right)^2\right],
\end{equation}
where $(u^\mu) = \gamma(1, v_x, v_y, v_z)$ is the four-velocity and $d \tau = d
t/\gamma$ is the differential proper time. The total emitted energy
within the interaction time $t_{int}$ is then given by $E_{rad} =
\int_0^{t_{int}} P d t$, which is the amount of energy that is
\textit{spontaneously scattered} by the particle according to
classical electrodynamics. In this picture, the scattered photons
occupy very different regions of the phase space (for both frequency
and direction) from the incoming photons, so we ignore stimulated
emission (which will be discussed in the next section). We have 
checked that the radiative back-reaction force is negligible compared
to the Lorentz force for the entire parameter space considered in
this paper. However, as we show below, the cumulative energy
loss could be significant such that a large fraction of the FRB wave
energy is scattered away. This in turn provides a constraint on the
gas density of the FRB environment.

\begin{figure*}
% \centerline{\hbox{\includegraphics[width=9cm, height=13cm,
%     angle=0]{constraint_ee.png}}}
% \centerline{\hbox{\includegraphics[width=9cm, height=13cm,
%     angle=0]{constraint_ep.png}}} 
\includegraphics[width=9cm, height=6cm, angle=0]{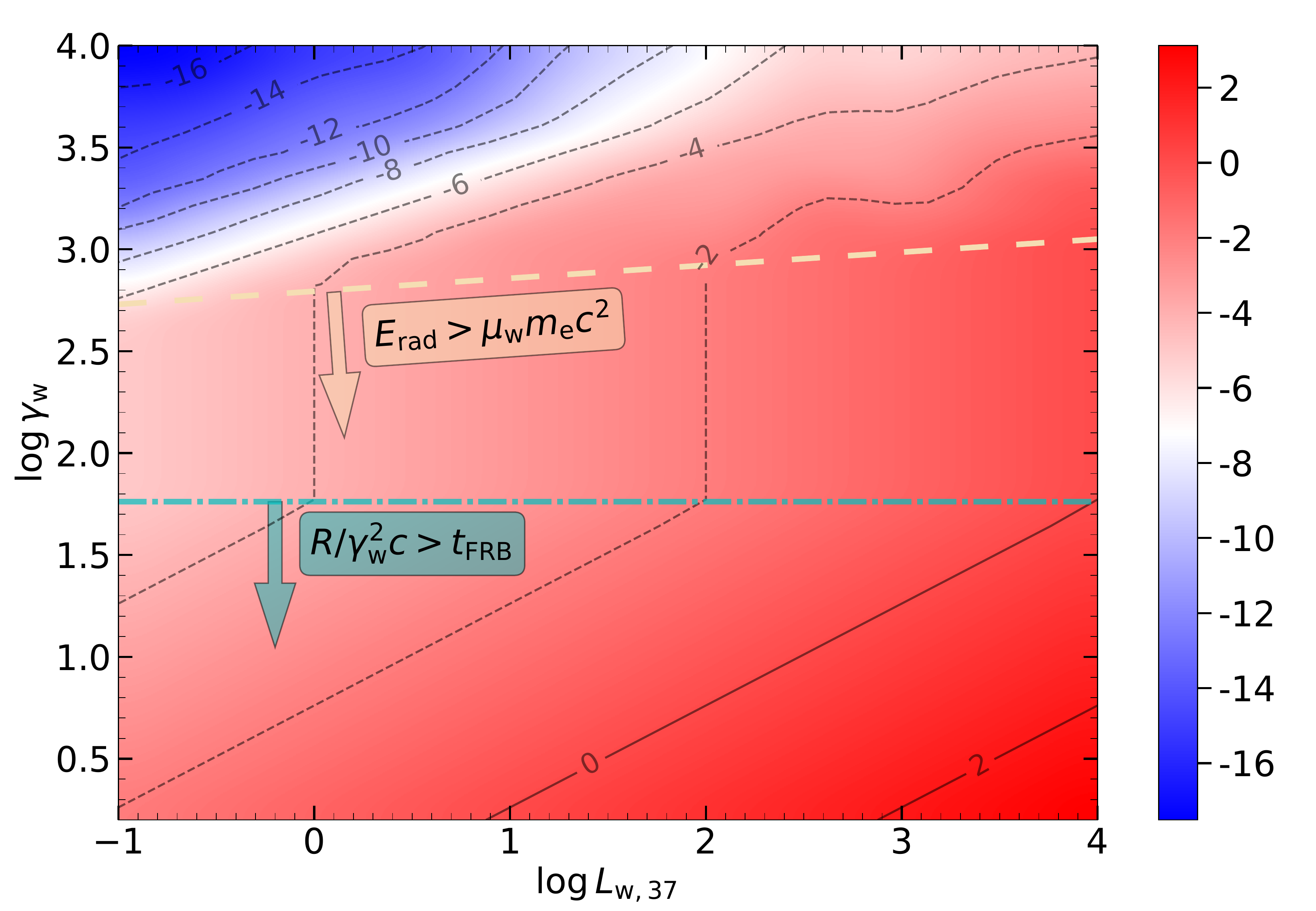}
\includegraphics[width=9cm, height=6cm, angle=0]{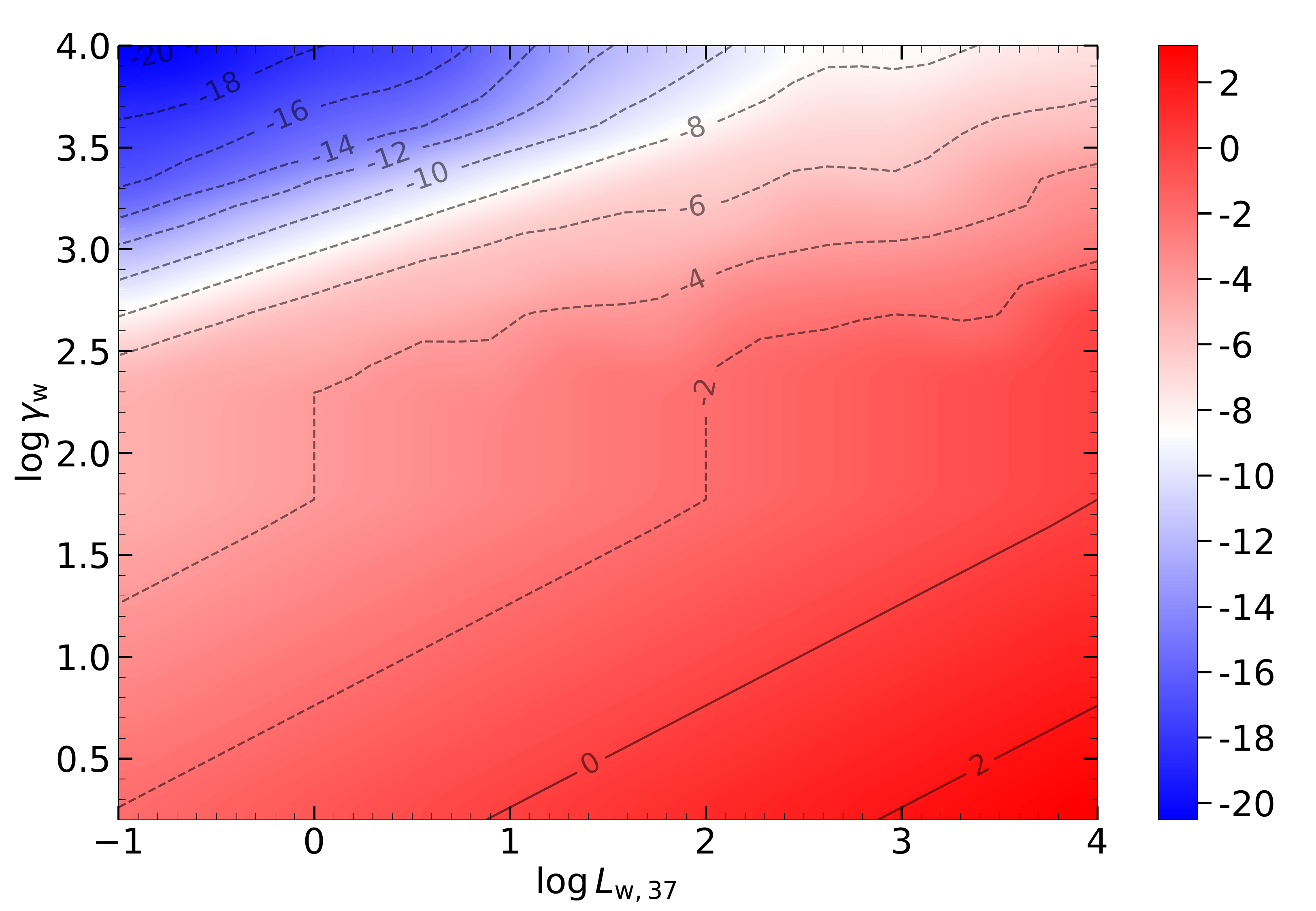}
% \vskip -0.7cm
\caption{The ratio between scattered energy and FRB energy
  $\mathrm{log}(E_{sca}/E_{FRB})$ as a function of wind power
  $L_w=10^{37}L_{w,37}\rm\, erg\,s^{-1}$ and
  LF $\gamma_w$. The region with
  $\mathrm{log}(E_{sca}/E_{FRB}) > 0$ is ruled out. We consider a
  typical weak burst from FRB 121102 with $t_{FRB}=0.1\rm\,ms$ and
  $L=10^{41}\rm\,erg\,s^{-1}$ at $\nu=1\rm\,GHz$. We assume the
  source to be located much below the 
  light cylinder of a neutron star ($\ll 10^{10}\,$cm) and take the
  characteristic FRB-wind interaction radius to be
  $R=10^{10}\,$cm. The left panel is for $\mu_w=1$ (corresponding to
  magnetized $e^{\pm}$ wind), and the right panel is for $\mu_w =
  10^{3}$ (magnetized electron-proton wind). The thick dashed and
  dash-dotted lines on the left panel mark the different regimes of
  the parameter space (see eq. \ref{eq:6}). Note that the particle 
  trajectory for $\omega_B \gg \omega$ and $a \gg 1$ is chaotic, and 
  we have carried out convergence tests with different time resolutions 
  to confirm the qualitative results.
}
\label{fig0}
\end{figure*}

Consider the FRB waves propagating through a strongly magnetized
relativistic wind with luminosity $L_w$ and LF
$\gamma_w$. An order unity fraction of the wind power is carried by
Poynting flux and a fraction $\mu_w^{-1}<1$ is carried by
electron (and positron) kinetic energy flux. At radius $R$ (much
greater than the light cylinder of a neutron star), the magnetic field
strength in the lab frame is $B \simeq \sqrt{ L_w/(r^2c)}$. For a
pulse of duration $t_{FRB}$, the interaction time between a particle
and the FRB wave is
\begin{equation}
  \label{eq:3}
  t_{int} \simeq \min(R/c,\ \gamma_w^2 t_{FRB}).
\end{equation}
We calculate the cumulative scattered energy for each electron in the
wind comoving frame $E_{rad}'$ and then Lorentz transform that to the lab
frame $E_{rad} = \gamma_wE_{rad}'$. The total number of electrons
participating in the interaction near characteristic radius $R$ is
\begin{equation}
  \label{eq:5}
  N_e \simeq {L_w \over \mu_w \gamma_w m c^2} \max\left({R\over
      \gamma_w^2 c}, \  t_{FRB}\right). 
\end{equation}
If the total scattered energy exceeds the total wind energy $\mu_w N_e
\gamma_w m c^2$, then the wind should be significantly accelerated by
the photon momentum which decreases $E_{rad}$.  Thus, the scattered
energy should generally be written as 
\begin{equation}
  \label{eq:4}
  E_{sca} = N_e \min( E_{rad},\ \mu_w\gamma_w m c^2).
\end{equation}
Therefore, we obtain the ratio between scattered energy
and the FRB wave energy
\begin{equation}
  \label{eq:6}
  \begin{split}
      {E_{sca}\over E_{FRB}} \simeq {L_{w,37} \over L_{43}}
    \min\left({E_{rad}\over \mu_w \gamma_w m c^2},
    1\right)  \max\left({R_{13}\over 3\gamma_w^2 t_{FRB,-3}}, 10^{-6}
  \right). 
  \end{split}
\end{equation}
Fig. \ref{fig0} shows $\mathrm{log}({E_{sca}/E_{FRB}})$ as a function of
$L_w$ and $\gamma_w$ for a typical weak burst from FRB 121102. If the
FRB source is within the magnetosphere of a neutron star (below the light
cylinder), we ruled out the presence of a mildly relativistic 
($\gamma_{w}\lesssim10$) wind from the progenitor star with 
$L_w\gtrsim 10^{39}\rm\,erg\,s^{-1}$, regardless of its
composition (pair or electron-proton), since the FRB pulse would 
lose a large fraction of its energy. We note that the persistent radio 
source associated with FRB 121102 (luminosity $\sim
10^{39}\rm\,erg\,s^{-1}$, at projected distance of $<40\,$pc,
Chatterjee et al. 2017, Marcote et al. 2017) may still be powered by an
ultra-relativistic $\gamma_w\gg 10$ wind like that in the Crab Nebula.

\section{Particle acceleration due to induced Compton scatterings}
\label{pmIC}

In this section, we calculate particle acceleration 
by scattering photons when the occupation numbers of both the initial and the
final photon states are very large, i.e. due to induced Compton (IC) 
scatterings. The calculations described here are valid when the wave 
nonlinearity parameter ``a'' (eq. \ref{aparam}) is less than 1; IC
scatterings for highly nonlinear waves is a complicated problem and
will be taken up in a future paper, however, nonlinear effects when 
$a>1$ are included in numerical calculations presented in Fig. \ref{fig2}.

It is best to view the process from the rest frame of an electron.
A photon of wave-vector $\bkp$ is scattered to $\bkps$ in electron rest frame.
The momentum kick given to the electron in this scattering is $\bkp - \bkps$.
The inverse of the process where $(k_2\hatbkps)\rightarrow\bkp$ gives an almost
exact opposite kick to the electron; $\hatbkps \equiv \bkps/|\bkps|$ is a
unit vector. The non-zero difference between the
two is due to the electron recoil in this scattering process so that 
$k_1' \not= k'$.

The rate of momentum transfer to an electron by a narrow beam of photons
with wave-vector within $d^3k'$ and its inverse process is
\begin{multline}
   {\bf \delta p'} = {c\hbar\over 2}{d\sigma_{_T}\over d\Omega_{\chi'}} 
   \bigg[(1 +\nkps) \nkp (\bkp - \bkps) {d^3k'\over 4\pi^3} + \\
    \quad\quad (1 +\nkp) n_{{\bf k_2}'}
    ({\bf k_2'} - \bkp) {d^3k'_2\over 4\pi^3} \bigg]d\Omega_{\chi'},
   \label{dpa}
\end{multline}
where 
\begin{multline}
    \cos\chi' \equiv \hatbkp\cdot\hatbkps, \; k_1' = k' - \Delta k', \;
       k_2' = k' + \Delta k', \\
     dk_2' = dk'(1 + 2\Delta k'/k'), \; {\rm and}\;\;
     \hbar\Delta k' = {\hbar^2 k'^2\over mc} (1 - \cos\chi')
   \label{k1k2}
\end{multline}
is the momentum recoil suffered by the electron due to scattering one
photon of momentum $\hbar\bkp$ by an angle $\chi'$. The factor $4\pi^3$
in the denominator in equation (\ref{dpa}) is because the number
of distinct photon quantum states in volume $d^3k'$ is $d^3k'/4\pi^3$.

We are considering the case where $\nk \gg 1$ is a function of $|\bk|$
within a cone, in the {\bf k}-space, of opening angle $\theta_s \ll 1$. 
Expressing $k_1'$ and $k_2'$ in terms of $k'$ (eq. \ref{k1k2}), and 
Taylor expansion of $\nkps$ \&
$n_{k_2'}$ in terms of $\nkp$ transforms equation (\ref{dpa}) to the following
expression
\begin{multline}
   {\bf \delta p'} = {d\sigma_{_T}\over d\Omega_{\chi'}} 
      {\hbar k'^3 dk' d\Omega_{\chi'} \over 4\pi^3}  \Bigg[ (\hatbkp -
     \hatbkps)\d\Omega_{\bf{k'}}  - (\hatbkp - \hatbkps)\d\Omega_{\bf{k_1'}} \\
    + {\Delta k'\over k'} (\hatbkps \d\Omega_{\bf{k'}} + 
     \hatbkps \d\Omega_{\bf{k'_1}}) \\ + \,{2(\bkps - \bkp)\Delta k'\over k'} 
  {\partial (\nkps k_1'^2)\over k_1'\nkps \partial k'_1 } \d\Omega_{\bf{k'_1}} 
    \bigg] c\nkp\nkps 
   \label{dpb}
\end{multline}

Integrating the above expression over incident and scattered photon directions
($\Omega_{\bf{k'}}$ \& $\Omega_{\bf{k_1'}}$) and frequency yields the total
rate of momentum deposit to the electron by the induced-Compton scatterings. 
The $\d\Omega_{\bf{k'}}$ and $\d\Omega_{\bf{k_1'}}$ integrals,
nested inside the square bracket in eq. (\ref{dpb}), are carried out subject 
to the condition that the scattering angle $\chi'$ is held fixed. Thus,
the first two terms in the square bracket cancel exactly when the angular 
integral over photon propagation direction is performed. The integral of 
the last term over $\Omega_{\bf{k_1'}}$ is smaller than the third term by 
at least a factor $\theta_s'^{-2}$ ($\theta_s'$ is half of angular size of the
radiation beam in the electron rest frame). This is because both the incident 
and scattered photons lie within the angle $\theta'_s$ for the 
induced-Compton scattering\footnote{Scattering a photon outside the 
radiation beam ($\chi' \gae \theta_s'$) is a weaker process than the induced 
Compton within the beam for the FRB parameters being considered here.} to be 
relevant and therefore, $|\bkps - \bkp| \lae k'\theta_s'^2/2$. Thus,
the last term in equation (\ref{dpb}) can be ignored as well. This 
simplifies the expression for the radiation force on the electron considerably
\begin{equation}
{\d{\bf p'}\over \d t'} \approx {c\hbar\over 2\pi^3} \int \d k'\, k'^2\int
   d\Omega_{\chi'} {d\sigma_{_T}\over d\Omega_{\chi'}} \int \d\Omega_{\bf{k'}}
\, \hatbkps \Delta k' \nkp\nkps.
   \label{dpdta} 
\end{equation}

Since $k' - k'_1 = \Delta k' \ll k'$ (see equation \ref{k1k2}), we can
take $\nkps\approx \nkp$, and the integral over $\Omega_{\bf{k'}}$ gives
$(\pi \theta_s'^2 \Delta k' \nkp^2) \hat{\bf z}$; the axis of the photon
beam cone is along $\hat{\bf z}$, and the photon occupation number ($\nkp$)
is taken to be angle-independent inside the radiation cone, i.e. for 
$\theta' < \theta_s'$.
Thus, equation (\ref{dpdta}) reduces to
\begin{multline}
   {\d{\bf p'}\over\d t'} \approx {3\hbar^2\sigma_{_T}\theta_s'^2\over 4\pi^2m} 
   \int \d k'\, k'^4 \nkp^2\int_{0}^{\theta'_s} d{\chi'} \sin\chi'\,
   \sin^2\Theta'\, \\
   \quad\quad\quad\quad \times\, \sin^2(\chi'/2) \, \hat{\bf z}, 
   \label{dpdtb}
\end{multline}
where we made use of equation (\ref{k1k2}) for $\Delta k'$ and 
\begin{equation}
    {d\sigma_{_T}\over d\cos\chi'} = {3\sigma_T\over 4} \sin^2\Theta',
\end{equation}
which is the differential scattering cross-section of an electron for linearly 
polarized radiation; where $\Theta'$ is the angle between the electric
vector of incident radiation and the momentum vector of scattered photons 
in the electron rest frame ($\Theta' \approx \pi/2$ for small angle 
scatterings that are relevant for IC acceleration).

The final two integrals are straightforward and are carried out assuming that
$\theta_s' \ll 1$ and $\nkp$ is a smooth function of $k'$:
\begin{equation}
  {\d{\bf p'}\over \d t'} \approx {3\hbar^2\sigma_{_T}\theta_s'^6 k'^5 \nkp^2
       \over 64\pi^2m}
\end{equation}
We can transform this equation to lab frame by noting that 
$dp'_z/dt' = dp_z/dt$ since the Lorentz boost is in the same direction
as $\delta{\bf p}'$ and the change to the particle energy in the comoving frame 
is proportional to $|\delta{\bf p}'|^2$. Moreover, 
\begin{multline}
  \quad\; \sin\theta_s' = {\sin\theta_s\over\gamma(1-\beta\cos\theta_s)}\approx
      2\gamma\theta_s, \; \quad {\rm for} \quad (\theta_s\gamma)\ll 1, \\
     \quad k' = k\gamma(1 - \beta\cos\theta_s)\approx k/2\gamma 
     \;\quad {\rm for} \quad (\theta_s\gamma)\ll 1, 
\end{multline}
which leads to
\begin{equation}
   {\d p_z\over \d t} \approx {3\hbar^2\sigma_{_T}\theta_s^6 k^5\gamma 
   \nkp^2 \over 32\pi^2 m}.
   \label{dpdtd}
\end{equation}

The photon occupation number, $\nkp$, is a Lorentz invariant quantity and
it can be easily shown to be
\begin{equation}
  n_{k} = {c^2 L_\nu \over 8\pi^2 \theta_s^2 R^2 h \nu^3} \sim
       {c^2 L \over 8\pi^2 \theta_s^2 R^2 h \nu^4}
   \label{nk}
\end{equation}
where $L_\nu$ is the specific luminosity (isotropic equivalent), $R$ is 
distance from the FRB source where particle acceleration is 
being considered. Substituting this into equation (\ref{dpdtd}) we finally 
obtain
\begin{equation}
   {\d p_z\over \d t} \approx {3\sigma_{_T}\theta_s^2 L^2\gamma 
   \over 256\pi^3 R^4\nu^3 mc}.
   \label{dpdtf}
\end{equation}

If the inertia of the medium is dominated by electrons
and positrons then $p_z = m c\beta\gamma$, and the equation 
for LF of the particle is
\begin{equation}
   {\d\beta\gamma\over \d t} \approx {3\sigma_{_T}\theta_s^2 
   L^2\gamma \over 256\pi^3 R^4\nu^3 m^2 c^2}.
   \label{dgamdt}
\end{equation}

This result is easy to understand as follows. Since for induced-Compton
scatterings, the scattered photon lies within the photon beam of
opening angle $\theta'_s$, the scattering cross-section is 
$\sigma_{_T} {\theta'_s}^2\, n_k$. In each scattering, the electron is 
recoiled and the momentum impulse it receives -- when we subtract the
contribution from its inverse process -- is $\sim h^2 
\nu'^2{\theta'_s}^2/(2 m c^3)$. The number of photons streaming outward
per unit area and time is, $L'/(4\pi R^2 h\nu')$. Combining all of these 
pieces we find the radiative force on an electron to be, $\sigma_{_T} L n_k  
h\nu \theta_s^4\gamma/(8\pi R^2 m c^3)$; where we made use of Lorentz 
transformations, i.e. $\theta'_s \sim \gamma \theta_s$, $\nu' \sim 
\nu/\gamma$, $L'\sim L/\gamma^2$. Substituting for $n_k$ from equation 
(\ref{nk}) we arrive at an expression for the radiative force on the
electron that is within a factor 2 of that given in equation (\ref{dpdtf}).
It should be noted that particle acceleration is more severe when the medium
through which the coherent radiation propagates is moving away from the source
at relativistic speeds.

\section{Constraints on FRB radiation mechanisms}
\label{appa}

We consider two cases in separate sub-sections. The first one is where the
magnetic field in the source is weak, $B\lae 10^3$G, so that the cyclotron
frequency is less than 1 GHz and the scattering of FRB radiation is
not affected by the magnetic field. The other case is that of a strong
magnetic field which is analyzed in \S\ref{appb}.

\subsection{The medium through which the FRB radiation is passing has weak 
   magnetic field (cyclotron frequency $\lae$ GHz)}
\label{appa1}

The FRB scenario considered in this section is where the coherent radiation 
is produced when a relativistic jet from a compact object interacts with 
the circum-stellar medium (CSM) and a fraction of the jet energy is 
converted to GHz photons. We provide general constraints on the
viability of this model. 

The ability of the shock model to reproduce the observed duration of FRBs,
and their frequency can be robustly constrained by combining a few
physical considerations. For much of the following discussions we will ignore
factors of order unity.

If the FRB radiation were to be produced at radius $\lae 10^{13}$cm, the 
nonlinear parameter associated with the radiation at this radius is 
$a\gae 5$ (eq. \ref{aparam}) and electrons upstream of the shock front are 
accelerated to LF $2a^2 \gae 50$ (see \S\ref{pmEMB}). This high 
LF of 
particles upstream completely changes the shock dynamics as well as the 
cyclotron/synchrotron maser instability growth rate that has been 
suggested for FRB radiation (Plotnikov \& Sironi 2019). Furthermore,
induced Compton scatterings 
upstream of the shock front provide important constraint on the shock
model for FRB radiation.

The induced-Compton (IC) scattering optical depth ahead of the shock front
is given by (e.g. Lyubarsky 2008, Lu \& Kumar 2018)
\begin{equation}
   \tau_{ic} \sim {\sigma_T L n_w (c t_{_{FRB}})\over 8\pi^2 R^2 m \nu^3} \sim
  {4 L_{43}\, n_{w,2} t_{_{FRB,-3}}\over R_{13}^2 \nu_9^3} 
  \label{tau-ic}
\end{equation}
This equation is valid 
for both IC scatterings within the photon beam and outside the photon
beam as long as $R \sim 2ct_{_{FRB}}/\theta_s^2$; where $\theta_s$ is the 
beam size of the coherent radiation at $R$ (the shock radius) -- 
$\theta_s \approx \gamma_{sw}^{-1}$ when the radiation is 
produced in a relativistic shock and the LF of the shocked fluid wrt
the upstream unshocked plasma is $\gamma_{sw}$. The IC optical depth 
given in equation (\ref{tau-ic}) is in the CSM rest frame. We need to modify 
this equation if the upstream CSM is a wind with LF $\gamma_w$.
The luminosity in the
wind rest frame $L' \approx L/\gamma_w^2$, $\nu' \approx \nu/\gamma_w$, and
the burst duration in the wind frame is $t'_{FRB}\approx \gamma_w^2\, t_{FRB}$.
Thus, the IC optical depth can be rewritten as
\begin{equation}
   \tau_{ic} \sim {\sigma_T L' n'_w (c t'_{_{FRB}})\over 8\pi^2 R^2 m \nu'^3} 
    \sim {\sigma_T L L_w \gamma_w (c t_{_{FRB}})\over 32\pi^3 m^2 c^3 
    \mu_w R^4 \nu^3},
\end{equation}
where $L_w$ and $\mu_w$ are wind luminosity and magnetization parameter, and we
used the relation $L_w = 4\pi R^2 n'_w m c^3 \mu_w \gamma_w^2$ to get rid
of $n'_w$. The requirement that $\tau_{ic} <1$
provides an upper bound on the CSM density 
\begin{equation}
    n_w(R) = n'_w\gamma_w \lae (30\,{\rm cm}^{-3}) { R_{13}^2 \nu_9^3 
    \over L_{43}\, t_{_{FRB,-3}}\gamma_w^2}.
  \label{n-max} 
\end{equation}

For density $n_w\ll 10$ cm$^{-3}$, the relativistic jet deceleration radius
and the place where coherent radiation is produced is $\gae 10^{13}$cm (see
\S\ref{maser-in-shock}).
This alleviates the problem associated with the acceleration of upstream 
$e^\pm$ away from the shock front to high LF due to the electric field of 
the coherent FRB radiation. Moreover,
at the larger radius, particle acceleration due to induced Compton 
scatterings also poses less of a problem as we describe next. However, 
the frequency of maser photons produced in the shock is below the
observing band for even CHIME at this low CSM density; this point is 
discussed further in \S\ref{maser-in-shock}.

The LF of plasma upstream of the shock front -- accelerated by
induced-Compton scatterings of FRB radiation -- can be calculated using
equation (\ref{dgamdt}), which is rewritten below for the relativistic case 
in a more convenient form
\begin{align}
   {d\gamma\over dt} &= {\gamma\over t_{acc}}, \quad\quad {\rm where} \\
     t_{acc} & \sim {256\pi^3 R^4\nu^3 m^2 c^2\over 3\sigma_{_T}\theta_s^2 L^2}
     \sim (3{\rm x10^{-13}s}) {R_{13}^4 \nu_9^3\over L_{43}^2 \theta_s^2}.
   \label{dgamdtf}
\end{align}
This shows that the LF of particles increases exponentially on a very short 
time scale, and they attain a terminal LF of $\sim 3\theta_s^{-1} \approx 
3 \gamma_{sw}$ when the radiation field in the particle comoving 
frame becomes nearly isotropic and little momentum is imparted to electrons 
in further scatterings. If the cold upstream medium moves away from the
source with LF $\gamma_w$ then $t_{acc}$ is smaller by a factor $\gamma_w$
in the lab frame.

The leading front of the radiation that is produced by the shock wave is 
ahead of the shock by a distance\footnote{This 
assumes that the wave frequency, $\omega$, is at least a factor few 
times $\gamma_{sw}$ larger than the plasma frequency in the upstream medium. 
This is the situation with many FRB radiation production scenarios including 
the cyclotron/synchrotron maser instability mechanism that operates in the 
shock transition layer and downstream of the shock front. If this condition 
were not satisfied then the FRB radiation would not be able to travel away 
from the shock front to be received by the observer.}
 $\sim \delta R/(2\gamma_{sw}^2)\sim c\, t_{_{FRB}} (\delta R/R)$; 
where $\delta R$ is the distance the shock has traveled since the
onset of the radiation. Thus, the time available in the lab frame for 
an upstream particle to be accelerated to $\gamma\sim \gamma_{sw}$ by the 
radiation front before it is swept up by the shock is roughly 
$t_{_{FRB}}/5$. This defines a radius, $R_{ra}$, beyond which 
radiative acceleration can be ignored. This radius is determined from 
the condition that $t_{acc}\sim t_{_{FRB}}/10$:
\begin{equation}
   R_{ra} \sim (5.5{\rm x}10^{13}\,{\rm cm}) \, t_{_{FRB,-3}}^{2/5} L_{43}^{2/5}
        \nu_9^{-3/5},
    \label{R-ra}
\end{equation}
where we took $\theta_s\sim \gamma_{sw}^{-1}$, and $\gamma_{sw}^2 \sim 
R_{ra}/(2c\,t_{_{FRB}})$ (see eq. \ref{tfrb}). We note that $R_{ra}$
depends extremely weakly on the effective particle inertia (per
electron) as $R_{ra}\propto m_{eff}^{-1/5}$. Therefore, $R_{ra}$ given
by eq. (\ref{R-ra}) does not change by more than a factor 4 even if 
the inertial of the upstream plasma is dominated by ions or magnetic 
field instead of $e^{\pm}$. We also note that $R_{ra}$ is larger
by a factor $\gamma_w^{3/5}$ when the CSM upstream of the coherent source 
is moving away with LF $\gamma_w$.

If FRB radiation is produced when the shock front is at $R\lae R_{ra}\sim 
10^{14}$cm, then the e$^\pm$ plasma upstream of the shock is accelerated 
to LF of order $3\gamma_{sw}$ due to the induced-Compton scattering of the 
radiation. Since the upstream particle velocity is larger than the shock 
front speed, and directed away from it, the effect of the IC scatterings is 
to prevent particles from approaching the front. Thus, radiation produced by
the shock for a time much less than $t_{_{FRB}}$, traveling upstream, 
accelerates plasma away from it, and halts further production of radiation.
IC accelerated particles slow down as they move outward and share their 
momenta with a larger number of particles. However, even the swept up medium 
has relativistic speed up to a distance $[\tau_{ic} \delta E/4\pi n 
m c^2]^{1/3}\sim 5{\rm x}10^{13}{\rm cm}\, 
[\tau_{ic}(\delta E)_{38}/n_2]^{1/3}$ from the shock front; where $\delta E$ 
is the energy of the radiation pulse produced by the shock before the process
is shut off, and $\tau_{ic}$ is the optical depth of the CSM to IC scatterings
out to the distance $5{\rm x}10^{13}$cm (given by eq. \ref{tau-ic}).
We see that just 1\% of a typical FRB radiation produced at $R\lae 10^{14}$cm 
can drive CMS to speed $\sim c$ and halt the radiation production process.

\subsubsection{Constraining maser mechanisms in shocks for FRB radiation}
\label{maser-in-shock}

Let us consider that the relativistic jet from a compact object 
interacts with a cold wind with the following properties: the 
wind is composed of particles of mass $m$, 
has LF $\gamma_w$, its magnetization parameter (the ratio
of magnetic to particle kinetic energy densities) is $\mu_w$, energy
density in wind comoving frame is $u'_w$, and its
luminosity in lab frame is $L_w \approx 4\pi R^2 u'_w \gamma_w^2 c$.
Similarly, the parameters of the relativistic jet are: $\gamma_j$, $\mu_j$,
$u'_j$ and $L_j$.
Let us take the LF of the shocked wind plasma wrt the unshocked wind to
be $\gamma_{sw}$, and the shocked jet plasma LF wrt the unshocked jet fluid 
is $\gamma_{sj}$. It can be shown using the continuity of mass, momentum
and energy flux across the shock front that for a relativistic, magnetized
plasma, the energy density of the shocked wind (in shocked fluid rest
frame) is $u'_{s,wind} \approx \gamma_{sw}^2 u'_w$, and similarly
$u'_{s,jet} \approx \gamma_{sj}^2 u'_j$ (e.g. Kennel \& Coroniti, 1984). 
The pressure equilibrium between 
the two shocked fluids, which are separated by the contact discontinuity 
surface, for highly relativistic systems, requires $u'_{s,jet} = u'_{s,wind}$. 
Moreover, the LF of the jet wrt the wind $\sim \gamma_j/\gamma_w$,
should be equal to $\sim \gamma_{sj}\gamma_{sw}$, i.e. two different ways
of calculating the relative speed of the jet and the wind should agree.
Combining these two relations we find
\begin{equation}
   \gamma_{sw}^4 \approx {\gamma_j^2 u'_j\over \gamma_w^2 u'_w} \quad {\rm
     or}\quad   \gamma_{sw} \approx \left[ {L_j\over L_w}\right]^{1/4}.
   \label{gam-sw}
\end{equation}

The deceleration radius of the jet is the distance it travels before 
half of its energy is transferred to the circum-stellar medium. The total
energy of the shocked wind, in the lab frame, is 
$\sim 4\pi R^3 u'_w \gamma^2_{sw}$. Thus, the deceleration radius is 
\begin{equation}
  E_j \equiv L_j t_{jet} \approx 8\pi R_{da}^3 u'_w \gamma^2_{sw} \implies
   R_{da} \approx {c E_j \gamma_w^2\over 2L_w \gamma^2_{sw} }, 
    \label{r-da1}
\end{equation}
where $t_{jet}$ is the time duration in the lab frame over which the
relativistic jet was launched from radius $R_{jet}$ in the magnetosphere
of the magnetar. The Shock LF after the deceleration radius is given by 
the energy conservation equation (\ref{r-da1}):
\begin{equation}
   \gamma_{sw}(R>R_{da}) \approx \gamma_{sw}(R_{da}) \left[ {R\over R_{da}}
         \right]^{1/2}.
\end{equation}

The FRB lasts for a time duration of order the jet deceleration time in the 
observer frame, $t_{da}^{obs}$, when roughly 50\% of the coherent radiation is 
generated. The observer frame deceleration time can be calculated from
$R_{da}$ and LF of the shock front in the lab frame\footnote{The Lorentz 
factor of the shock front is larger than the LF of the shocked plasma 
($\gamma_{sw}$) by a factor $\sim \mu_w^{1/2}$ (e.g. Kennel \& Coroniti, 
1984).}, 
$\gamma_{sw}\mu_w^{1/2}\gamma_w$, using the standard formula for a 
source that is moving toward the observer with relativistic speed, e.g.
Kumar \& Zhang (2015)
\begin{equation}
  t_{da}^{obs} \approx {R_{da}\over 2c \left[\gamma_{sw}\gamma_w\mu_w^{1/2}
    \right]^2} \approx {t_{jet}\over 4\mu_w},
   \label{tda}
\end{equation}
where we used equations (\ref{gam-sw}) and (\ref{r-da1}) to arrive at the
second equality in (\ref{tda}). 
The jet launching time, $t_{jet}$, ought to be smaller than a few ms because 
of FRB energetics\footnote{The total energy of the relativistic jet
(isotropic equivalent) for a typical FRB is of order $10^{42}$erg if
the radiation production efficiency is a few percent. Considering
the ms duration of FRBs, the jet energy has to come from the magnetic field
and not the neutron star rotation; the rate of energy extraction for the latter
is set by the dipole radiation formula and is far smaller than what we see
for FRBs (see eq. \ref{Lw-NS}). For a magnetar with surface magnetic field
of $10^{15}$G, the total magnetic energy above the radius $R$ is 
2x10$^{42} R_8^{-3}$erg. Therefore, the jet has be launched at a radius
no larger than $\sim 10^8$cm, i.e. $R_{jet} \lae 10^8$cm and $t_{jet}\approx
  R_{jet}/c \lae $3 ms.}. Thus, the shock model gives burst duration (eq.
\ref{tda}) much smaller than the observed few ms width of FRBs if the 
magnetization parameter of the wind into which the relativistic jet is 
running into ($\mu_w$) is larger than order unity.
The possibility that the FRB radiation might be produced
at a radius much larger than $R_{da}$ will be considered later in this
sub-section; the constraint on $\mu_w$ when $R\gg R_{da}$ turns out
to be similar. It should also be noted that upstream particles are
accelerated by IC scatterings more rapidly and out to larger distances
when $\gamma_w \gg 1$ as discussed in \S\ref{appa1}, and that is an
additional constraint on $\gamma_w$ and $\mu_w$.

Substituting for $\gamma_{sw}$ from equation (\ref{tda}) into (\ref{r-da1}), 
and expanding $L_w \approx 4\pi R_{da}^2 \mu_w m c^3 \langle
   n'_w\rangle\gamma_w^2/3$, we find
\begin{equation}
   R_{da} \sim \left[{ 3 E_j t_{_{FRB}}\gamma_w^2\over 4\pi m c \langle 
   n'_w\rangle }\right]^{{1\over4}} \sim (5{\rm x10^{12} cm}) 
   \gamma_w^{{1\over2}} \left[ {E_{42} t_{_{FRB,-3}}\over \langle 
    n'_w\rangle_{_4}}\right]^{{1\over4}},
  \label{R-shock}
\end{equation}
where $\langle n'_w\rangle$ is the mean CSM particle number density in
its rest frame within $R_{da}$; $\langle n'_w\rangle= 3 n'_w(R_{da})$
for a steady wind CSM. 
This expression is independent of the LF of the relativistic jet, and
the magnetization of the wind. The second 
part of equation (\ref{R-shock}) is
obtained by taking the average particle mass in the wind to be the electron
mass; $R_{da}$ is smaller for an ionic wind by a factor 7. 
Considering the weak dependence of $R_{da}$ on burst energy and CSM density, 
it is hard for the FRB source, according to the shock model scenario, to be 
at a radius very different from 10$^{13}$cm. However, upstream particle
acceleration by the emergent radiation renders the shock model not viable
at $R\lae 10^{13}$cm. 

\begin{figure*}
  \includegraphics[width=\textwidth,height=12.0cm]{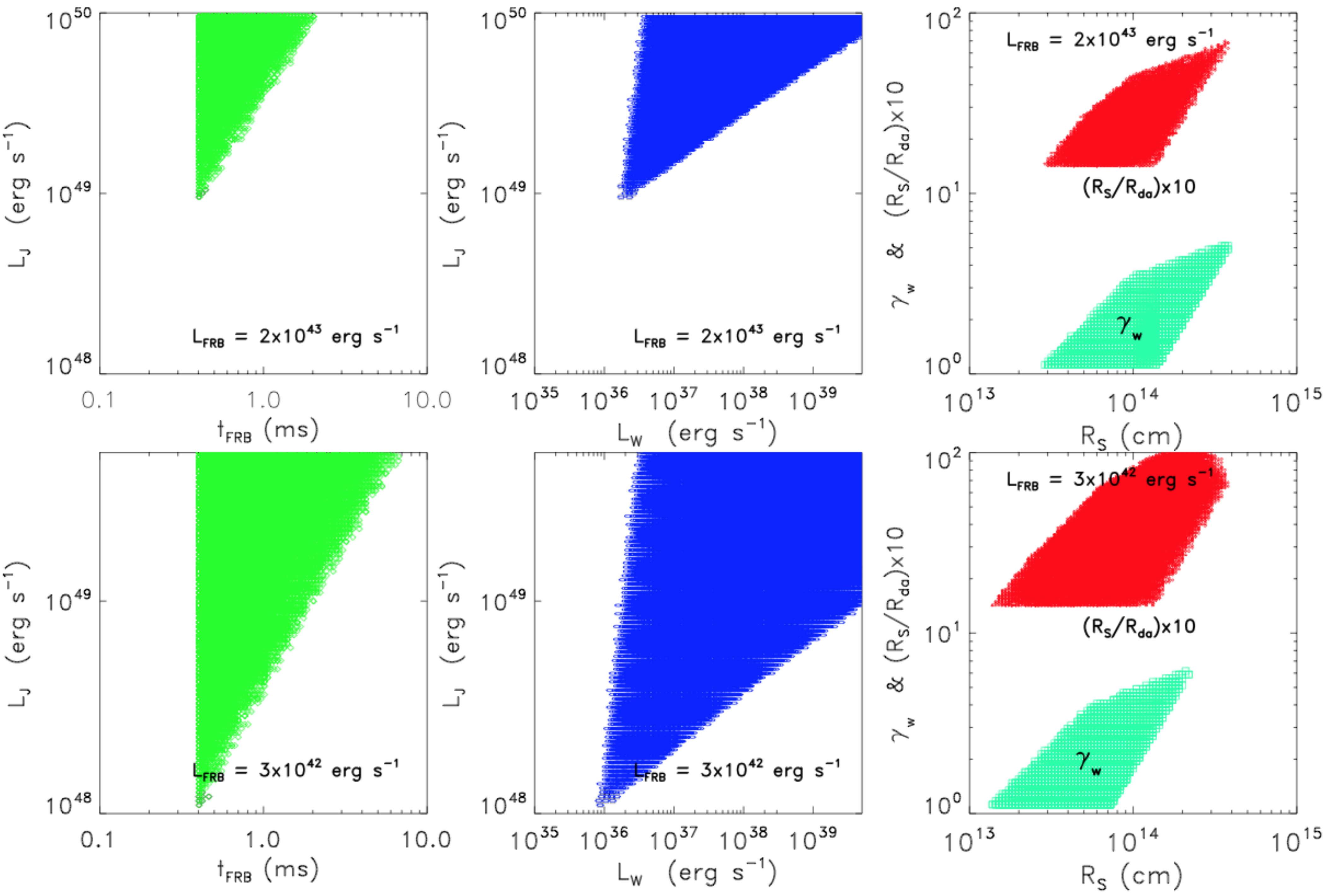}
   \vskip -0.2cm
\caption{We show results of numerical calculations for FRB emission 
according to maser mechanisms operating in shocks that result from a
relativistic jet of luminosity $L_j$ colliding with a cold wind 
of luminosity $L_w$, both of which are produced by the same compact object.
Each panel shows a pair of parameters that survive the constraint
imposed by particle acceleration upstream of the shock front that 
shuts off the generation of coherent radiation. Constraints are 
placed on the particle acceleration time due to induced-Compton (IC)
scatterings -- $t_{acc}(R_s) > t_{_{FRB}}/10$; where $t_{_{FRB}}$ is 
burst duration in observer frame which is calculated using eq. \ref{tfrb}, 
and $t_{acc}(R_s)$, given by eq. \ref{dgamdtf}, is particle acceleration 
time due to IC at the shock radius $R_s$ (we note that our numerical 
calculation of $t_{acc}$ includes nonlinear effects when the wave 
nonlinear parameter $a>1$ whereas 
eq. \ref{dgamdtf} is valid only for $a\ll1$). For the purpose of calculating
upstream particle accelerations by the IC process, we have taken the
observed, isotropic, FRB luminosity to be either 2x10$^{43}$ erg s$^{-1}$
(the top three panels) -- the median luminosity of ``non-repeating FRBs''
is $\sim 10^{44}$ erg s$^{-1}$ ( e.g. Luo et al. 2018; Ravi 2019) --
or 3x10$^{42}$ erg s$^{-1}$ (bottom three panels), which is roughly the
median luminosity for the bursts of the repeater FRB 121102
(e.g. Michilli et al. 2018; Hessels et al. 2019). We do not impose
a prior on this luminosity as it is dependent on the details of the
maser process whereas the focus of this paper is to provide general
constraints that should apply to all maser mechanisms operating 
in shocks. We emphasize that the luminosity information is used only 
for the calculation of upstream particle acceleration and nothing else. 
We show in this figure the parameter space which yields 
the burst duration $t_{_{FRB}} > 0.4$ ms, 
and the peak frequency $\nu_B$ (given by eq. \ref{nuB}) greater than 
0.4 GHz (which is the lowest frequency at which these bursts
have been detected by CHIME -- Amiri et al. 2019a, 2019b). The parameter
search is cut off at $L_j > 10^{50}$ erg s$^{-1}$ and $L_w > 5$x10$^{39}$
erg s$^{-1}$. The luminosity for the biggest magnetar flare we have
ever observed (SGR 1806–20) was about 2x10$^{47}$erg s$^{-1}$ in 
$\gamma$-rays (Hurley et al. 2005; Palmer et al. 2005), and the 
energy was smaller in relativistic outflows by a factor $\sim 10^2$
(Gelfand et al. 2005, Granot et al. 2006). The upper limit on $L_j$ 
of 10$^{50}$ erg s$^{-1}$ we have set in our numerical calculations is 
a factor $\sim 10^5$ larger than the luminosity of mildly-relativistic 
outflows for SGR 1806–20.
The observed FRB frequency is taken to be the cyclotron frequency $\nu_B$.
   What we find is that the maser-in-shock model for FRB radiation requires
   the isotropic luminosity of the relativistic jet responsible for the 
   FRB radiation to be 
   $\gae 10^{49}$ erg s$^{-1}$ (top and bottom left and middle panels), and that 
   means that the total outburst energy of the repeater FRB 121102 in
   just one year exceeds the magnetic field energy of a magnetar with surface
   field strength of $\sim 10^{15}$G; we should point out that the high 
   value of $L_j$ is dictated entirely by the requirement that the emission 
   is produced at a sufficiently large radius $R_s$ and shock LF $\gamma_{sw}$
   (eq. \ref{gam-sw}), so as to avoid excessive particle acceleration upstream 
   of the shock front due to IC, and at the same time produce burst duration 
   $\gae 0.4$ ms; no constraint is placed as to how the maser mechanism operates and 
   its efficiency. The top and bottom left panels show that the jet energy increases
   very rapidly with increasing burst duration.
   The top and bottom right panels show that the 
   observed radiation is produced at a radius that is a few times 
   larger than the deceleration radius of the jet, and the wind 
   LF is between $\sim$1 and 4 for reasons that are explained in 
   \S\ref{maser-in-shock}.
   The upstream induced-Compton optical depth is included in
   these calculations, but it turns out to be a substantially weaker 
   constraint than IC acceleration. The last point to note is that the 
   maser-in-shocks model has no solutions for FRB luminosity 
   $\gae 10^{44}$ erg s$^{-1}$; there are solutions if we allow $L_j$
   to be larger than $\sim10^{51}$ erg s$^{-1}$, but that poses
   problems for the total energetics.
    }
  \label{fig2}
\end{figure*}

As argued before, the magnetization parameter of the wind CSM cannot be larger
than order unity otherwise the FRB duration, according to the shock model, 
would be much smaller than the observed duration. Thus, we take $\mu_w\sim 1$.
Moreover, the asymptotic LF of a magnetized wind satisfies the 
relation $\gamma_w \sim
\mu_w^{1/2}$ (Goldreich \& Julian, 1969; Granot et al. 2011). So we
take $\gamma_w$ to be of order unity as well.
Thus, the expected CSM particle 
density at $R_{da}\sim 10^{13}$cm, for an electron-positron wind, is
\begin{equation}
    n_w \sim {L_w\over 4\pi R^2 m c^3} \sim (3{\rm x}10^5\, {\rm cm}^{-3})
            L_{w,37} R_{13}^{-2},
    \label{n-ext}
\end{equation}
where our choice of wind luminosity of 10$^{37}$ erg s$^{-1}$ is for a 
young magnetar (see eq. \ref{Lw-NS}). The high particle density upstream of 
the shock front makes the CSM highly opaque to induced-Compton scattering 
(eq. \ref{tau-ic}), and that is another problem with the shock model for
FRB radiation.

FRB radiation could be generated in shocks at $R\gae 10^{14}$cm, where 
upstream particles are not forced by the radiation to move at relativistic 
speed away from the shock front. However, if we demand
that $R\sim R_{da}$, so that the FRB radiation is produced efficiently, 
then that requires the energy of the relativistic jet to be $\gae 10^{46}$erg 
or $n \lae 1$ cm$^{-3}$ (see eq. \ref{R-shock}). The large energy 
requirement seems problematic especially for the repeater FRB 121102; 
the total energy of outbursts in the last 10 years for this object exceeds
the available energy in magnetic fields of even an extreme magnetar if
each outburst has $\gae 10^{46}$erg energy. On the other hand, if we take 
the CSM density to be sufficiently small so that $R_{da} \gae 10^{14}$cm then 
the frequency of coherent radio emission produced in shocks is much smaller 
than the observed frequencies for FRBs (this point is amplified below).

We next explore the possibility that the FRB radiation is generated at a
radius $R_s \gae 10^{14}$cm, and relax the efficiency considerations so that 
$R_s$ can be much larger than the deceleration radius ($R_{da}$) of the 
relativistic jet.

The characteristic photon frequency for many maser instabilities in plasmas
is of order the plasma or cyclotron frequency depending on the
nature of the instability (the latter is $\mu_w^{1/2}$ 
times the former; $\mu_w\equiv B^2/(8\pi n_w m c^2)$ is plasma magnetization 
parameter). The cyclotron frequency in the shocked wind frame is
\begin{equation}
   \nu_B' \approx {q B_w'\over 2\pi m c},
\end{equation}
and in the observer frame it is
\begin{equation}
   \nu_B \approx \nu_B' \gamma_{sw} \gamma_w \sim {q L_w^{1/2}\gamma_{sw}\over
      2\pi m R_s c^{3/2}}.
\end{equation}
We can eliminate $\gamma_{sw}$ using the FRB duration for the shock model, viz.
\begin{equation}
  t_{_{FRB}} \approx {R_s\over 2c (\gamma_{sw}\gamma_w\mu_w^{1/2})^2},
   \label{tfrb}
\end{equation}
where the factor $\mu_w$ in the denominator is because the shock front LF
wrt to the unshocked wind is $\gamma_{sw}\mu_w^{1/2}$ (the LF of shocked
plasma wrt the unshocked wind is $\gamma_{sw}$). Thus, we arrive at the
following expression for the cyclotron frequency of the shocked wind in the
observer frame (which should be of order the FRB frequency for any maser 
mechanism apart for a possible factor $\sim\mu_w^{1/2}$):
\begin{equation}
  \nu_B \approx {q L_w^{1/2}t_{_{FRB}}^{-1/2}\over 2^{3/2}\pi m c^2 R_s^{1/2} 
    \gamma_w\mu_w^{1/2}} \approx (7{\rm x10^8Hz}) {L_{w,37}^{1/2}
   t_{_{FRB,-3}}^{-1/2}\over R_{s,14}^{1/2} \gamma_w\mu_w^{1/2} }.
   \label{nuB}
\end{equation}
This frequency is at the lower end of the radio band at which FRBs are observed.
We see from the above equation that there is little room for 
$\gamma_w\mu_w^{1/2}$ to be much larger than order unity unless 
$L_w \gg 10^{37}$ erg s$^{-1}$. 

The dipole wind luminosity for a 30 year old magnetar with surface field 
strength of 10$^{15}$G is (e.g. Goldreich \& Julian, 1969)
\begin{equation}
   L_w \sim 10^{37} {\rm erg\, s}^{-1} \; B_{NS,15}^{-2} t_9^{-2},
   \label{Lw-NS}
\end{equation}
as long as $t$ is larger than the spin-down time $t_{sd}$, which is given by
\begin{equation}
    t_{sd} \approx 500\, {\rm s}\; B_{NS,15}^{-2} P_{-3}^2,
   \label{tsd}
\end{equation}
where $P_{-3}$ is the pulsar rotation period in unit of 10$^{-3}$s. The
luminosity is roughly constant for $t \lae t_{sd}$.

The maximum dipole wind luminosity at time $t$ after the birth of a NS is
\begin{equation}
   L_w^{max} \sim 10^{39} {\rm erg\, s}^{-1} \; P_{-1}^{-2} \, t_9^{-1},
  \label{Lw-max}
\end{equation}
which corresponds to $t_{sd}\sim t$, and the surface magnetic field of
\begin{equation}
    B \sim (7{\rm x10^{13} G)}\, P_{-1} t_9^{-1/2}.
\end{equation}
where $P_{-1}$ is the pulsar rotation period in unit of 0.1 s at time $t$.

It is entirely possible that the pulsar magnetic field, especially for a young
system, is highly non-dipolar. Moreover, the magnetar wind might not be 
rotationally powered, but instead launched by magnetic field dissipation. 
For these cases, the wind luminosity provided by equations 
\ref{Lw-NS}--\ref{Lw-max} does not apply, and for that reason we 
consider $L_w$ as high as 10$^{40}$ erg s$^{-1}$ in all of our 
numerical calculations presented in Fig. \ref{fig2}.

The electron density associated with the magnetar $e^\pm$ wind at 
$R_s$ is $n_w \sim L_w/(4\pi R_s^2 m c^3\gamma_w \mu_w)\sim 
3{\rm x}10^3$cm$^{-3} L_{w,37} R_{s,14}^{-2} (\gamma_w\mu_w)^{-1}$. The 
density marginally exceeds the upper limit given in equation (\ref{n-max}) 
-- to avoid the medium upstream of the shock-front to become opaque 
to induced Compton scatterings -- unless $\gamma_w\mu_w \gg 1$.

The LF of the shock at $R_s \gae 10^{13}$cm should be $\gae 10^3 
\gamma_w^{-1} \mu_w^{-1/2}$ in order to produce a ms duration burst
(eq \ref{tfrb}). This requires the luminosity of the
relativistic jet, obtained from equation (\ref{gam-sw}), to be $L_j\sim
L_w \gamma_{sw}^4 \gae 10^{49} \gamma_w^{-4}\mu_w^{-2} L_{w,37}$ 
erg s$^{-1}$. 

The allowed parameter space for the maser-in-shocks model of FRB is shown in 
Figure \ref{fig2}. The parameters are for a FRB with observed luminosity
of 2x10$^{43}$erg s$^{-1}$ (top three panels), which corresponds to the
low end of the luminosity of non-repeaters (e.g. Luo et al. 2018; Ravi 2019).  
And the lower three panels of Figure \ref{fig2} show results for radio
luminosity of 3x10$^{42}$erg s$^{-1}$ which is the median luminosity for bursts
of the repeater FRB 121102 (e.g. Michilli et al. 2018; Hessels et al. 2019).
The observed luminosity is used for the calculation of particle acceleration 
upstream of the shock front, and for no other aspect of the maser mechanism.
As the left top and bottom panels of the figure show, the energy requirement
for maser-in-shock model grows very rapidly with increasing $t_{FRB}$. For 
bursts of duration longer than $\sim2$ ms, $L_j \gae 10^{49}$ erg s$^{-1}$.
The energy requirement is reduced if the FRB frequency is much larger than 
the cyclotron frequency $\nu_B$ considered in these calculations.

There are no solutions when the relativistic jet luminosity $L_j \lae 10^{49}$ 
erg s$^{-1}$ (top left and middle panels of Fig. \ref{fig2}). And that is a
 serious problem for the maser-in-shock model for FRBs. 
The well studied repeater FRB 121102 has been observed for about 10 years,
and $\gae 10^2$ outbursts have been detected during the small observing 
time invested to following this object. 
The object has had numerous outbursts with radio luminosity is GHz band 
of $\gae 10^{43}$ erg s$^{-1}$. Each of these bursts require, according 
to the maser model, $L_j \gae 10^{49}$ erg s$^{-1}$, and therefore, 
the total energy needed for outbursts in a year is at least $10^{48}$ erg 
if the efficiency for converting the magnetic energy to relativistic outflows 
is 100\%. We know empirically that 
giant magnetar outbursts convert less than a few percent of 
magnetic energy to mildly-relativistic outflows\footnote{The giant magnetar 
outburst of December 27, 2004 (SGR 1806–20) released 4x10$^{46}$ erg in 
$\gamma$-rays (Hurley et al. 2005, Palmer et al. 2005) whereas the energy 
in mildly relativistic outflow, from radio observations, was estimated to be a 
few times $10^{44}$erg (Gaensler et al. 2005, Gelfand et al. 2005, 
Granot et al. 2006). The peak 
luminosity for this burst, which was larger by a factor $\sim 10^2$ 
compared with the previous two most luminous SGRs, was $\sim$2x10$^{47}$ 
erg s$^{-1}$. The duration of the initial $\gamma$-ray spike, which carried 
most of the energy release, for all three giant SGR flares was 0.2s.},
and perhaps a much smaller fraction to ultra-relativistic jet that is 
invoked by maser-in-shocks models for FRB radio emission.
This greatly exacerbates the energy problem -- 
the energy requirement for maser-in-shocks model to support the activities
of the FRB repeater for one year is $\gae 10^{50}$erg and that is
hard for magnetic fields, even those as large as 10$^{16}$G, to provide. 
We note that the value of $L_j$ in our calculations is dictated solely by 
the shock LF $\gamma_{sw}$ and $R_s$ such that the observed duration of bursts 
comes out to be of order a few ms, and not by any efficiency considerations.

The maser model yields no solution for FRB
luminosity $\gae 10^{44}$erg s$^{-1}$ when the acceleration of 
upstream particles by IC scatterings shuts off the maser mechanism in 
less than 1 ms unless we consider the luminosity of the relativistic
jet $L_j \gae 10^{51}$ erg s$^{-1}$; $L_j$ scales linearly with the
FRB radio luminosity. We note that the analysis of ASKAP sample of bursts
shows that there are FRBs with luminosity as high as  
$\sim 10^{46}$ erg s$^{-1}$ (Lu \& Piro, 2019).

The main result of this sub-section is that it is highly unlikely that
the FRB radiation is produced at a distance much larger than a 
few hundred neutron star radii. Particles in the region $10^9 \lae 
R_s \lae 10^{13}$cm suffer very strong radiative acceleration which 
disrupts the photon generation process. The energy requirement for
maser-in-shock models operating at $R_s \gae 10^{13}$cm is very challenging
($\gae 10^{48}$ erg in ultra-relativistic outflows in one year 
for repeaters such as FRB 121102).
For $R \lae 10^9$cm, the strong magnetic field of a magnetar suppresses the
induced-Compton scatterings and acceleration of particles by
the electric field of FRB coherent radiation (discussed in the next 
sub-section). Therefore, the generation of coherent photons can proceed 
unimpeded close to the neutron star.

Although, almost all the discussions in this section have 
explicitly considered the scenario where FRB radiation is generated in 
shocks, the same physical considerations --- 
acceleration of particles to high LF by the emergent radiation in the
vicinity of the source region --- apply to any other model for FRBs such
as plasma maser type mechanism that operate far away from the neutron
star surface and outside the light cylinder.

\subsection{FRB radiation generation and propagation in a region of strong 
magnetic field}
\label{appb}

Photon-electron scattering cross-section is significantly modified in 
the presence of strong magnetic fields when the cyclotron frequency
$\nu_B\gae\nu$. The cross-section for an X-mode photon\footnote{An 
X-mode is a linearly polarized EM wave with the direction of the wave electric
field vector perpendicular to the plane of static magnetic field and the
wave-vector.} is (e.g. Canuto et al. 1971) 
\begin{equation}
   \sigma_{\rm x} = {\sigma_{_T}\over 2} \left[ {\nu^2\over (\nu+\nu_B)^2} +
     {\nu^2\over (\nu-\nu_B)^2}\right],
   \label{sigmax}
\end{equation}
and the cross-section when the wave electric field is not perpendicular
to the static magnetic field is:
\begin{equation}
   \sigma_\parallel = \sigma_{_T} \left[ \sin^2\theta_{kB} + {\cos^2\theta_{kB}\over2}
     \left\{{\nu^2\over (\nu+\nu_B)^2} + {\nu^2\over (\nu-\nu_B)^2}\right\}
     \right],
   \label{sigmao}
\end{equation}
where the cyclotron frequency
\begin{equation}
    \nu_B = {q B\over 2\pi m c},
\end{equation}
$\sin\theta_{kB}\approx\theta_{kB}$ is the dot product of unit vectors 
along the static magnetic field and wave electric field as measured 
in the electron rest frame, and $\nu$
is EM wave frequency also in the electron rest frame. These formulae for
the cross-section apply only when the EM wave nonlinearity parameter 
$a_\parallel\equiv qE_\parallel/mc\omega$ (see eq. \ref{aparam})
is much less than 1; $E_\parallel$ is the component of wave electric 
vector amplitude along the static magnetic field. For the X-mode we are
considering here, the angle between the wave electric field and the magnetic
field is very close to $\pi/2$ (eq. \ref{thetaEB}), and $a_\parallel \ll 1$;
this point is further addressed below (\ref{thetaEB}).

We see from equation (\ref{dgamdtf}) that for $R \lae 10^{13}$cm the 
timescale for particle acceleration due to induced-Compton scatterings 
are extremely short, provided that the cross-section for scattering a
photon by an electron into an unoccupied state is not drastically smaller
than $\sigma_{_T}$. However, as we see from equation (\ref{sigmax}),
$\sigma_{\rm x}$ is smaller than $\sigma_{_T}$ by a factor 
$(\nu_B/\nu)^2$ in the region of high magnetic field. Attenuation of 
O-modes\footnote{An O-mode is a linearly polarized EM wave with the 
direction of the wave electric
field vector in the plane of static magnetic field and the
wave-vector.}
by interactions with electrons in the medium in the vicinity of the
source is also suppressed by a factor $\sim\theta_{kB}^{-2}$;
modes with $\theta_{kB}\not\ll 1$ are unlikely to 
be able to escape the immediate vicinity of the source region intact. 

Consider an X-mode generated at radius $R_s$ (measured from the center 
of the host neutron star with strong magnetic field) such that $R_s \lae
10^{7}$cm.
It was shown by Lu et al. (2019) that the electric-vector of an X-mode 
traveling through a medium with non-uniform magnetic field rotates in
such a way as to keep the wave-electric field pointing nearly perpendicular 
to the local magnetic field and the wave-vector as long as
the plasma density is sufficiently large;
the index of refraction of the medium for this mode is very close to unity, 
so the wave-vector does not rotate.
The RMS angle $|{\bf \hat E_w\times \hat B}|$ as the wave
travels away from the source region is given by (Lu et al., 2019)
\begin{equation}
    \theta_{kB} \approx {2\pi c\,\nu \over R_B\omega_p^2}
   \label{thetaEBa}
\end{equation}
where $\omega_p^2 = 4\pi q^2 n/m$ is the plasma frequency, $n$ is the electron
density and $R_B$ is the radius of curvature of magnetic field lines at the
current location of the wave at $R$; $R_B\sim R/\theta$ at polar coordinate 
($R, \theta$) wrt the magnetic axis. Let us take the $e^\pm$ density 
at $R$ to be ${\cal M}$ 
times the Goldreich-Julian density (Goldreich \& Julian, 1968)
\begin{multline}
    n = {\cal M}\, \nGJ = { {\cal M}\, {\bf B\cdot\Omega_{ns}}\over 2\pi qc} 
    \approx {{\cal M}\, B_{ns}\Omega_{ns}\over 2\pi qc}\left( {R_{ns}\over R}
    \right)^3 \\
    \approx 10^{13}{\rm cm}^{-3} {\cal M}\, B_{ns,15} \Omega_{ns} 
     \left( {R_{ns}\over R} \right)^3,
     \label{nR}
\end{multline}
where $\Omega_{ns}$ is the angular velocity of the NS.

Substituting this into (\ref{thetaEBa}) we find 
\begin{equation}
  \theta_{kB} \sim 6{\rm x}10^{-11} {\cal M}^{-1} \nu_9 R_{B,8}^{-1} B_{NS,15}^{-1} \Omega_{ns}^{-1} (R/R_{ns})^3,
    \label{thetaEB}
\end{equation}
inside the freeze-out radius given by equation (\ref{Rfo}).
The wave nonlinearity parameter along the magnetic field $a_\parallel \equiv 
q E_\parallel/(mc\omega) = q E \theta_{kB}/(mc\omega) \ll 1$ even out to 
$R/R_{ns}\sim 10^3$ or $R \sim 10^9$ cm, if ${\cal M} \sim 10^3$ ($R_B \sim
R/\theta \sim 10^2 R$ in the magnetic polar-cap region). Therefore, 
it is appropriate to use equation (\ref{sigmao}) for X-mode scattering cross
section even for bright FRBs.

The rotation of wave electric field ceases at a radius, called the
freeze-out radius, where the plasma density becomes too small to be able 
to provide the current needed to rotate the wave electric vector. This 
occurs at a radius where (Lu et al., 2019)
\begin{equation}
  {\omega_p^2\over \omega^2} \sim {ac\over \omega R_B} \implies
{R_{fo} \over R_{ns}} \sim 2 {({\cal M} B_{ns,15}\Omega_{ns})^{{1\over2}}
       \over L_{43}^{-{1\over4}} } \left[{ R_B\over R_{ns}}\right]^{{1\over2}};
   \label{Rfo}
\end{equation}
$R_B$ in this equation is calculated at $R_{fo}$.
The FRB radiation might be produced along open magnetic field lines in the
polar cap region which has an angular size of $\theta_{pc} = 
[\Omega_{ns}R_{ns}/c]^{1/2} \sim 5.8{\rm x}10^{-3} (R_{ns,6} 
\Omega_{ns})^{1/2}$ rad. 
Taking $\theta\sim \theta_{pc}$ and substituting $R_B\sim  R_{fo}/\theta$ 
in equation (\ref{Rfo}) leads to
\begin{equation}
   {R_{fo} \over R_{ns}} \sim 400\, {\cal M} B_{ns,15} 
          (\Omega_{ns}/R_{ns,6})^{1/2} L_{43}^{-1/2}.
 \label{Rfoa}
\end{equation}
The angle $\theta_{kB}$ at the freeze-out radius is:
\begin{equation}
   \theta_{kB}(R_{fo}) \sim 10^{-5}\, {\cal M}\, B_{NS,15}
      L_{43}^{-1} \Omega_{ns}^{1/2} R_{ns,6}^{-3/2} \nu_9.
\end{equation}

Making use of equation (\ref{thetaEB}) we find that the photon-electron 
scattering cross-section, in the strong magnetic field regime 
($\omega_B\gg \omega$), when the wave electric field is not exactly 
perpendicular to the large scale magnetic field is
\begin{equation}
  \sigma_\parallel \sim \sigma_{_T} \theta_{kB}^2 \sim 2{\rm x}10^{-45}
    {\rm cm}^2 {\nu_9^2\over {\cal M}^{2} B_{ns,15}^{2}\Omega_{ns}^{2} 
       R_{B,8}^{2}} \left( {R\over R_{ns} }\right)^6,
   \label{sigmaparallel}
\end{equation}
as long as $\theta_{kB} > \nu/\nu_B$, which is the only case we are 
considering here.

The induced-Compton scattering optical depth in this case is
\begin{equation}
   \tau_{ic} \approx {3\sigma_\parallel n L \theta_s^2\over 64\pi^2 R\nu^3 m},
\end{equation}
where $\theta_s =\min\{ \ell_s/R, \gamma_s^{-1}\}$ is the angular size of the 
photon beam at $R$, $\ell_s$ is the transverse size of the FRB source 
from which photons are received at $R$, and $\gamma_s$ is the LF of the
FRB source.  Making use of equations (\ref{nR}) \& (\ref{sigmaparallel}) 
we find the optical depth to IC for $R\lae R_{fo}$
\begin{equation}
    \tau_{ic} \sim 10^{-7} {L_{43} \ell_{s,4}^2\over \nu_9 R_{B,9}^2 {\cal M}_3
       B_{ns,15}\Omega_{ns,1}}. 
\end{equation}
The IC optical depth is less than unity at all radii in the NS magnetosphere
at least out to the light cylinder.

Next, we look into particle acceleration and energy loss in the source region 
of the FRB pulse. In a strong magnetic field region, particle
acceleration time for $e^\pm$ plasma due to IC scattering is a slightly 
modified form of equation
(\ref{dgamdtf}) 
\begin{equation}
  t_{acc} \sim {256\pi^3 R^4\nu^3 m^2  c^2\over 3\sigma_{\parallel}
    \theta_s^2 L^2}. 
\end{equation}
Using equation (\ref{sigmaparallel}) leads to the following expression
for the acceleration time
\begin{equation}
   t_{acc} \sim (10^{-5}{\rm s}) {\nu_9 (B_{ns,15}\Omega_{ns,1} {\cal M}_3
       R_{B,9})^2\over L_{43}^2 \ell_{s,4}^2} \quad {\rm for} \quad R\lae 
    R_{fo}.
   \label{tacc_strongB}
\end{equation}
According to the coherent curvature model of FRBs (Kumar et al. 2017; 
Lu \& Kumar 2018), ${\cal M}\gae 10^3$ and $\ell_s\sim 10^4$cm, 
and so the IC acceleration time within the source region of FRBs is 
$\sim 10^{-5}$s. This is longer 
than $\ell_s/c$, the residency time of e$^\pm$ in the source region,
and hence the induced-Compton cannot adversely affect the generation
of FRB coherent radiation. In fact, our estimate of $t_{acc}$ inside the
source region is probably too small. This is because the electric field 
of the radiation is exactly perpendicular to the static magnetic field
(X-mode polarization) within the source and $e^\pm$s are stuck in the
lowest Landau level and have weaker than classically expected interaction 
with X-mode photons. 
It is also the case that the coherent curvature radiation requires a 
strong electric field along the static magnetic field. Particle 
acceleration time due to this electric field is of order 10$^{-16}$s 
(Kumar et al.  2017), and the electric force on $e^\pm$ far exceeds 
any non-zero IC scattering force.

Particles outside the source region, at larger radii, are accelerated to 
LF $\sim \theta_s^{-1}\sim R/\ell_s\sim 10^3 R_7/\ell_{s,4}$ when $t_{acc} 
\ll t_{_{FRB}}$; the FRB radiation becomes roughly isotropic in the particle 
rest frame when the LF approaches this value, and IC scattering force drops 
to zero. 
The total number of $e^\pm$ in the magnetosphere of a magnetar of spin period 
1 s is $\sim 4\pi {\cal M}\, \nGJ R^3 \sim 3{\rm x}10^{36} {\cal M}_3 
B_{ns,15}$. Therefore, the energy lost by the FRB radiation as it travels 
through the NS magnetosphere is $\lae 2{\rm x}10^{33}{\cal M}_3$ erg 
(isotropic equivalent). This energy loss is a tiny fraction 
of the total energy of the FRB coherent radiation\footnote{One might worry 
that electrons and positrons accelerated
to high LF could trigger a pair production avalanche, which
can sap the energy from FRB radiation. However, e$^\pm$ accelerated
by FRB radiation are moving away from the neutron star, and thus see the NS
surface emission highly red-shifted and not capable of pair production.
Moreover, the photon density associated with pulsar nebula emission is 
$\lae 10$ cm$^{-3}$, which is too small for launching pair cascade.}. 

The nonlinearity parameter for the FRB coherent radiation (described in 
\S\ref{pmEMB}) is $\sim 10^5$ at $R=10^8$cm. In the absence of the 
magnetic field of the magnetar, $e^\pm$ exposed to this radiation
would be accelerated to LF $\sim 10^{10}$. However, the strong magnetic
field of a magnetar suppresses particle acceleration drastically, 
and the LF e$^\pm$s attain at $R=10^8$cm is
close to unity\footnote{The angle between the electric field of radiation
and the local magnetic field direction minus $\pi/2$, defined to be
 $\theta_{kB}$, is given by eq. \ref{thetaEB}.
Thus, the wave nonlinearity parameter along the magnetic field
$a_\parallel \equiv q E \theta_{kB}/(mc\omega) \ll 1$ for $R\lae10^8$cm.
And, therefore, e$^\pm$ are accelerated along the magnetic field by the 
electric field of the FRB radiation to speeds much smaller than $c$ 
throughout most of the NS magnetosphere. The acceleration of particles
perpendicular to the static magnetic field when the cyclotron frequency 
($\omega_B$) is much larger than wave frequency $\omega$ is determined by 
a modified non-linearity parameter $a_\perp \equiv q E/(m c \omega_B)$, 
which is less than 1 for $R \lae 10^8$cm. Therefore, the particle speed 
perpendicular to the magnetic field is also sub-relativistic even quite 
far from the NS surface; at $R=10^9$cm, $a_\perp \sim 10$ and e$^\pm$ LF 
is $\sim10^2$.}. 
Therefore, particle acceleration by the
electric field of FRB radiation does not change the conclusion we
arrived at regarding the loss of FRB energy as photons travel
through the NS magnetosphere to arrive at Earth, i.e. the loss of
energy is negligible.

The main result of this subsection is that a FRB source located within 
a few 10s of neutron star radii of a magnetar can withstand the enormous 
radiation forces. Moreover, little energy is lost as the radiation travels
though the NS magnetosphere. The reason for this is entirely due to the 
strong magnetic field of a magnetar which suppresses scattering cross-section
and the efficiency of particle acceleration, i.e. in the absence of 
a strong magnetic field a large fraction of FRB radiation energy would
be imparted to particles along its path as the radiation travels away from 
the source.

\section{Conclusions}

We have investigated the effects of an intense FRB coherent 
radiation on plasma around the region where the radiation is produced. 
The purpose is to constrain source properties and the radiation mechanism for 
FRBs by using some fairly general physics considerations, and determine
conditions that a successful model should satisfy.
In particular, we calculate particle speed due to electric field of the 
radiation and the highly enhanced scattering (the induced-Compton scattering) 
by electrons and positrons 
of the coherent FRB radiation where each quantum state has an occupancy 
number of order 10$^{35}$. We find that electrons and protons are
accelerated to very high LF due to these forces. This
severely restricts some, otherwise, promising models for FRBs. One such
class of models invokes relativistic shocks and maser type instability 
operating in the shock transition zone or downstream of the shock front. 

We have shown that the coherent radiation 
traveling upstream of the shock front stirs up the plasma violently
-- $e^{\pm}$ have LF greater than 30 due to the strong electric field of the 
radiation -- even when the shock front is at a distance of 
10$^{13}$cm from the FRB compact progenitor star (see \S2). 
Furthermore, induced-Compton scatterings push the upstream plasma away
with a large LF as long as the shock front radius
is less than $\sim 10^{13}$cm, thereby preventing particles from 
approaching and crossing the shock front to keep the generation of 
GHz radiation going. 

At larger distances ($\gae 10^{13}$cm) these forces are fairly tame. However,
maser-in-shock models require an excessively large amount of energy 
($\gae 10^{46}$ erg) in ultra relativistic outflows to produce a 
burst at a few GHz frequency with luminosity $\gae 10^{43}$ erg s$^{-1}$, 
at these large distances (see \S\ref{maser-in-shock}); as a point of 
reference, the total energy in relativistic outflows in the biggest magnetar
flare ever observed (SGR 1806–20) was estimated to be $\sim10^{44}$ erg
from late time radio observations (Gelfand et al. 2005; Granot et al. 2006).
The total 
energy in relativistic jets for bursts produced by the repeater FRB 121102 in 
one year, for maser models, is required to be substantially larger than 
10$^{48}$ erg. This exceeds the energy 
in magnetic fields of a NS with strength 10$^{15}$G, and that is assuming an 
efficiency of 100\% for converting magnetic energy to relativistic jets; 
the actual efficiency is perhaps no larger than 1\% (\S\ref{maser-in-shock}).

If FRBs were to be associated with SGRs then that has consequences for
the maser-in-shock models. The $\gamma$-ray photons from the SGR will
be scattered by electrons upstream of the shock front heating them up. 
Whether the maser instability can survive this interaction is unclear. Let
us consider that the SGR associated with a FRB released 10$^{47}$ erg 
in $\gamma$-rays, and the $\gamma$-ray luminosity was 10$^{48}$erg $s^{-1}$;
the duration of the $\gamma$-ray pulse for SGR 1806–20 was about 0.1 s. The
$\gamma$-ray pulse is ahead of the shock front, when it is at radius $R$,
by the distance $\delta R\approx R/(2\gamma{sw}^2)\sim c t_{FRB}$. Upstream 
electrons will be heated to the temperature of $\gamma$-rays on a time scale 
of $t_{ic\gamma} \sim (1\, {\rm ms}) R_{13}^2/L_{\gamma,48}$. Considering 
this short time scale, and small Larmor radius ($\lae 5{\rm x}10^3$cm), 
the upstream particles would have anisotropic velocity distribution as they
enter the shock front, and that could affect the instability.

Stimulated Raman and Brillouin scatterings can also attenuate the
FRB radiation. The stimulated Raman process has been calculated by 
many people e.g. Gangadhara \& Krishan (1992), Thompson et al. (1994), 
Levinson \& Blandford (1995). Lyubarsky \& Ostrovska (2016) considered Raman 
scatterings in the context of stellar coronae model for FRBs and 
concluded that it does not provide more severe constraint on 
the propagation of FRB radiation than the induced-Compton scatterings. 
However, implications of stimulated Raman and Brillouin scatterings 
for the more recent models for FRBs needs to be investigated.

We have shown in \S\ref{appb} that the FRB radiation source operating is 
a region of very strong magnetic field (where the cyclotron frequency is 
much larger than the FRB radio wave frequency) alleviates these problems; 
the radiation forces are weaker in the presence of strong magnetic field,
and the plasma in the FRB source region is not dispersed quickly.
One of the reasons
for this is that the interaction cross-section between $e^\pm$ and X-mode 
photons is highly reduced in the presence of a strong magnetic field.
The coherent curvature (``antenna'') model for FRBs requires very 
strong magnetic field for its successful operation (Kumar et al. 2017; 
Lu \& Kumar 2018). It is shown in \S\ref{appb} that the source survives 
the radiative forces, and radiation traveling through the NS magnetosphere
suffers negligible loss of energy.

\section{acknowledgments}

We thank the referee, Roger Blandford, for helpful comments and suggestions.
WL acknowledges useful discussions with Lorenzo Sironi and Brian
Metzger on the physics of synchrotron instability, with Ben
Margalit on wind nebulae near magnetars, and we thank Bing Zhang for comments
and suggestions. WL is supported by the David
and Ellen Lee Fellowship at Caltech.

\end{document}